\documentclass[journal]{IEEEtran}
\usepackage{amsmath}
\usepackage{amsmath,amssymb}
\usepackage{stmaryrd}
\usepackage{graphicx,graphics,color,psfrag}
\usepackage{cite,balance}
 \usepackage{subfigure} 
\usepackage{caption}   
\allowdisplaybreaks
\usepackage{algorithm}
\usepackage{bm}
\usepackage{eufrak}
\usepackage{url}
\usepackage[english]{babel}
\usepackage{algpseudocode}
\usepackage{multirow}
\usepackage{enumerate}
\usepackage{cases}

\usepackage[utf8]{inputenc}
\usepackage[english]{babel}
\newtheorem{lemma}{Lemma}
 
\newtheorem{theorem}{Theorem}

\begin{document}
 \pdfminorversion=4
\graphicspath{{figure/}} 
\title{A Covariance-based User Activity Detection and Channel Estimation Approach with Novel Pilot Design \thanks{The work was supported in part by the Key Area R\&D Program of Guangdong Province with grant No. 2018B030338001, by the National Key R\&D Program of China with grant No. 2018YFB1800800, by Natural Science Foundation of China with grant NSFC-61629101, by Guangdong Zhujiang Project No. 2017ZT07X152, by Guangdong Basic and Applied Basic Research Foundation (Grant No. 2019A1515111140), by Shenzhen Peacock Plan under Grant KQTD2015033114415450, and by the Open Research Fund from Shenzhen Research Institute of Big Data under Grant No. 2019ORF01012.}  \thanks{Lei Cheng is with Shenzhen Research Institute of Big Data (SRIBD) (leicheng@sribd.cn). Liang Liu is with EIE Department, The Hong Kong Polytechnic University (liang-eie.liu@polyu.edu.hk). Shuguang Cui is with The Chinese University of Hong Kong, Shenzhen and SRIBD (shuguangcui@cuhk.edu.cn).}}

\author{Lei Cheng, Liang Liu and Shuguang Cui
}

\maketitle

\pagestyle{empty}  
\thispagestyle{empty} 

\begin{abstract}
This paper studies the massive machine-type communications (mMTC) for the future Internet of Things (IoT) applications, where a large number of IoT devices exist in the network and a random subset of them become active at each time instant. Building upon the fact that the covariance matrix of the received signal can be accurately estimated in the spatial domain if the base station (BS) is equipped with a massive number of antennas, we propose a covariance-based device activity detection and channel estimation strategy in a massive MIMO (multiple-input multiple-output)  aided mMTC system. For this strategy, a novel approach for the pilot sequence design is first provided, where the pilot of each device is merely determined by a unique phase parameter. Then, by estimating the phase parameters of the active pilot sequences that contribute to the received covariance matrix, an efficient algorithm is proposed to detect the active devices without the prior information about the total number of active devices. At last, given the estimation of active devices, channel estimation is conducted based on the conventional minimum mean-squared error (MMSE) approach. It is worth noting that our proposed strategy is able to obtain all the results in closed-forms, and is thus of much lower complexity compared to the existing strategies that are based on iterative algorithms for device detection and channel estimation.
\end{abstract}

\vspace{-12pt}
\section{Introduction}

To embrace the forthcoming era of Internet of Things (IoT), 3GPP has defined massive machine-type communications (mMTC) as a main use case for the fifth-generation (5G) cellular networks \cite{B1}. In a typical massive IoT connectivity scenario, although there exist a large number of IoT devices, merely a random subset of them are active at each time instant due to their sporadic data traffic. To facilitate mMTC, the so-called grant-free random access has been studied \cite{SPM}, in which the base station (BS) needs to detect the active devices and estimate their channels based on its received pilot sequences before decoding their messages. 

In the literature, the strategies for device activity detection and channel estimation in mMTC can be classified into two categories: the compressed sensing based strategy and the covariance-based strategy. For the former strategy, device activity detection and channel estimation is cast into a compressed sensing problem based on sparsity in user activity \cite{Liang_TSP, CS1}. Along this line, \cite{Liang_TSP} proposed to apply the approximate message passing (AMP) algorithm \cite{AMP} in mMTC, and it was shown that in the asymptotic regime when the number of antennas at the BS goes to infinity, the probabilities of missed detection and false alarm both go down to zero exponentially. Moreover, other compressed sensing algorithms were also applied in the mMTC systems \cite{CS1}. For the latter strategy, the key observation is that given the active devices, the covariance matrix of the received signal can be accurately estimated in the spatial domain if the number of antennas at the BS is large. In the massive MIMO (multiple-input multiple-output) system, a stochastic maximum likelihood (ML) method was proposed in \cite{Con1, Con2} to detect the active users based on the received covariance matrix. However, the user pilot sequences are randomly generated in the above strategies, which may not satisfy the constant power constraint of each pilot symbol in practice. Moreover, the above approaches all require iterative algorithms for device detection, which are time consuming when the number of users is large in the system. 

To tackle the above challenges, in this paper, we provide a novel pilot sequence design approach, based on which a closed-form covariance-based strategy is then proposed for device activity detection and channel estimation in a massive MIMO aided mMTC system. Specifically, each user is assigned with a unique phase parameter, and it generates its pilot sequence as a function of this parameter. It is shown that under our proposed pilot design, each received covariance matrix merely corresponds to one unique device activity pattern. Moreover, a closed-form algorithm is proposed to detect the active devices, which does not require the prior information about the total number of active users and only involves the eigenvalue decomposition (EVD) of some matrices associated with the received covariance matrix. At last, the minimum mean-squared error (MMSE) channel estimation is conducted given the estimated active users. 
\vspace{-1.5 em}
\section{System Model}
\vspace{-.8em}
Consider the uplink communications in a massive MIMO system, where the BS is equipped with a large number of antennas, and each user is equipped with single antenna. Let $M$ and $N$ denote the number of antennas at the BS and the number of users, respectively.  A block-fading channel model is adopted in this paper. We consider the Rayleigh fading channel, and the channel from the $n^{th}$ user to the BS is denoted by  $\boldsymbol h_n \sim \mathcal{CN}(\boldsymbol 0,\beta_n \boldsymbol I )$, where $\beta_n$ is the pass loss and  assumed to be known by the BS. Due to the sporadic nature of wireless traffic in mMTC, only $K \ll N$ users are active in each coherent block. The indicator variable $\lambda_n$ is introduced to represent the  activity pattern of the $n^{th}$ user as follows:
\begin{align}
 \lambda_n =\left\{
\begin{aligned}
& 1, ~ \mathrm{if~user}~n~\mathrm{is~active},  \\
& 0, ~\mathrm{otherwise}, 
\end{aligned}
~~~ \forall n.
\right.
\end{align}
Define $\mathcal{K}= \{k: \lambda_k=1\}$ as the set of active users.

In this paper, we consider the grant-free random access scheme \cite{SPM}, where the user activity $\lambda_n$'s and user channels $\boldsymbol h_n$'s have to be estimated in the first phase, and the user message is decoded in the second phase. Particularly, we focus on user activity detection and channel estimation in this work.  In this phase, the $n^{th}$ user is assigned a unique pilot sequence $\boldsymbol a_n=[a_n(1),...,a_n(L)]^T$ with length $L$, which is assumed to be smaller than the coherence length. Due to the fact that the number of potential users $N$ is usually extremely large in the context of mMTC, the length of each pilot sequence $L$ is typically less than the total number of users $N$, resulting in non-orthogonal pilot sequences $\{ \boldsymbol a_n \}_{n=1}^N$. 

At the beginning of each coherence block, the active users transmit their pilot sequences to the BS.  The received measurements at the BS can be modeled as 
\begin{align}
\boldsymbol Y = \sum_{n=1}^N \lambda_n \boldsymbol a_n \boldsymbol h_n^T + \boldsymbol Z,
\end{align}
where $\boldsymbol Z  = [\boldsymbol z_1, ... , \boldsymbol z_M]$, with $\boldsymbol z_m \sim \mathcal {CN}( \boldsymbol 0 , \sigma^2 \boldsymbol I )$, denotes  the additive white Gaussian noise (AWGN) at the BS. By defining  $\boldsymbol x_n \triangleq \lambda_n \boldsymbol h_n, n=1,...,N,$ as each user's effective channel, $\boldsymbol X=[ \boldsymbol x_1, ..., \boldsymbol x_N]^T$ as the collection of all users' effective channels, and $\boldsymbol A=[ \boldsymbol a_1,..., \boldsymbol a_N]$  as the collection of all users' pilot sequences,  the measurement model can be equivalently expressed as 
\begin{align}
\boldsymbol Y &= \sum_{n=1}^N \boldsymbol a_n \boldsymbol x_n^T + \boldsymbol Z = \boldsymbol A \boldsymbol X + \boldsymbol Z.
\end{align}

\vspace{-5pt}

Notice that detecting the active users and estimating their channels by recovering $\boldsymbol X$ based on (3) can be viewed as a compressed sensing problem, since $\boldsymbol X$ is a row-sparse matrix. From this perspective, research work \cite{Liang_TSP} used the AMP algorithm to estimate the row-sparse $\boldsymbol X$. However, the AMP-based method in \cite{Liang_TSP}  requires the prior information of the number of active users $K$, which is hard to get in practice. Moreover, for the AMP-based method \cite{Liang_TSP}, the user pilot sequences in $\boldsymbol A$ are i.i.d. generated, and it is hard to design a better pilot matrix $\boldsymbol A$ to improve the performance.

In this work, instead of the compressed sensing strategy, we employ a  covariance-based scheme to detect the active users and then estimate their channels. Specifically, given the user activity $\lambda_n$'s, the covariance matrix of the received signal is
\begin{align}
\boldsymbol \Sigma_{\boldsymbol Y}(\lambda_1,...,\lambda_N)= \mathbb E \left[\boldsymbol Y \boldsymbol Y^H \right]= \boldsymbol A \boldsymbol \Gamma  \boldsymbol A^H+\sigma^2 \boldsymbol I,
\end{align}
where $\boldsymbol \Gamma = \mathrm{diag}\{ \lambda_1 \beta_1,  ..., \lambda_N\beta_N\}$.  If the covariance matrix $\boldsymbol \Sigma_{\boldsymbol Y}(\lambda_1,...,\lambda_N)$ is known, it is possible to estimate $\lambda_n$'s based on (4). Then, after $\lambda_n$'s are estimated, the user channels can be recovered based on the received signal model (3). 

In practice, the covariance matrix in general has to be estimated over a long time duration given any user activity pattern. Since the user activity can change very fast in mMTC, there is no sufficient time to estimate $\boldsymbol \Sigma_{\boldsymbol Y}(\lambda_1,...,\lambda_N)$ accurately. However, our considered massive MIMO setup enables an alternative manner to estimate the covariance matrix $\boldsymbol \Sigma_{\boldsymbol Y}(\lambda_1,...,\lambda_N)$ in the spatial domain based on the following relationship:
\vspace{-5pt}
\begin{align}
\hat{\boldsymbol \Sigma}_{\boldsymbol Y}(\lambda_1,...,\lambda_N)=\frac{1}{M} \boldsymbol Y  \boldsymbol Y^H \approx \boldsymbol \Sigma_{\boldsymbol Y}(\lambda_1,...,\lambda_N).  
\vspace{-5pt}
\end{align}
Moreover, as  $M \rightarrow \infty$, it can be shown that $\hat{\boldsymbol \Sigma}_{\boldsymbol Y}(\lambda_1,...,\lambda_N)  \rightarrow  \boldsymbol \Sigma_{\boldsymbol Y}(\lambda_1,...,\lambda_N).$

In the following, we first focus on the ideal case when $M$ goes to infinity such that the estimation of $\boldsymbol \Sigma_{\boldsymbol Y}(\lambda_1,...,\lambda_N)$  is perfect. In this case, we will introduce our proposed scheme that can construct $\boldsymbol A$ effectively, detect user activity perfectly with low complexity, and estimate the channels accurately. Next, we will discuss about how to apply our scheme in a practical massive MIMO system when the BS is equipped with tens or hundreds of antennas such that the estimation of $\boldsymbol \Sigma_{\boldsymbol Y}(\lambda_1,...,\lambda_N)$ is accurate but not perfect.

\vspace{-11pt}
\section{Algorithm Design When $M \rightarrow \infty$}
We start to introduce our proposed user activity detection and channel estimation scheme by considering the asymptotic regime of $M \rightarrow \infty$ such that $\boldsymbol \Sigma_{\boldsymbol Y}(\lambda_1,...,\lambda_N)$  is estimated perfectly by (5), i.e., 
$\hat{\boldsymbol \Sigma}_{\boldsymbol Y}(\lambda_1,...,\lambda_N)  =\boldsymbol \Sigma_{\boldsymbol Y}(\lambda_1,...,\lambda_N)$. Particularly, we consider the case of 
\vspace{-5pt}
\begin{align}
L > K.
\vspace{-5pt}
\end{align}
In other words, the pilot sequence length is larger than the number of active users. This is a valid assumption since even with known user activity, channel estimation requires $L>K$  to be true in general to reduce the estimation error.

%
%
%
%
%
%
\vspace{-12pt}
\subsection{Identifiable Constant-Modulus Pilot Design}
In this subsection, a novel pilot construction scheme is presented. For the pilot vector of the $n^{th}$ user, it is designed to be 
\begin{align}
\boldsymbol a(\phi_n) = & \sqrt{p}[1, \exp(-j2\pi \delta \cos \phi_n ),  \exp(-j4\pi \delta \cos \phi_n), \nonumber \\
& ... ,  \exp(-j(L-1)2\pi \delta \cos \phi_n )]^T,
\end{align}
where $p$ denotes the identical transmit power of all the users,  $\phi_n \in [0, \pi]$, which can be set to be $\phi_n = n\pi/N, n = 1, ...N,$ in practice, and  $\delta \leq 1/2$. Then, define $\boldsymbol A (\boldsymbol \phi) = [\boldsymbol a(\phi_1), $ $\boldsymbol a(\phi_2), ..., \boldsymbol a(\phi_N) ]$ as the collection of all users' pilot sequences, which is a function of  $\boldsymbol \phi=[\phi_1,...,\phi_N]^T$. For this pilot design, it is observed that each pilot symbol is of constant power, which is desirable in practice such that the existing modulation schemes can be directly applied. At last, our pilot design also satisfies the following property.

\begin{theorem}
Given a pilot matrix $\boldsymbol A (\boldsymbol \phi) = [\boldsymbol a(\phi_1), $ $\boldsymbol a(\phi_2), ..., \boldsymbol a(\phi_N) ] \in \mathbb C^{L \times N} (L < N)$ with each column $\boldsymbol a(\phi_n)$ generated by (7), any $K$ columns in $\boldsymbol A (\boldsymbol \phi)$ are linearly independent. Moreover, different user activity patterns will yield different receive  covariance matrices, i.e.,  $\hat{\boldsymbol \Sigma}_{\boldsymbol Y}(\bar{\lambda}_1,...,\bar{\lambda}_K) \neq \hat{\boldsymbol  \Sigma}_{\boldsymbol Y}(\tilde{\lambda}_1,...,\tilde{\lambda}_K)$, if $\exists k, \bar{\lambda}_k  \neq  \tilde{\lambda}_k$.
\end{theorem}

A rigorous proof of Theorem 1 is given in  [8, page 272]. Note that each pilot sequence is an identifier for each user since different users are assigned with different pilot sequences. As a result, {Theorem 1} indicates that the perfect covariance matrix estimation $\hat{\boldsymbol \Sigma}_{\boldsymbol Y}(\lambda_1,...,\lambda_N)$ is sufficient for recovering the user pilot sequences and thus detecting the active users without error under our pilot sequence design given in (7). In the following, we show how to map an arbitrary $\hat{\boldsymbol \Sigma}_{\boldsymbol Y}(\lambda_1,...,\lambda_N)$ to the corresponding pilot sequences that contribute to it.

%
%
%
%
%
\vspace{-12pt}
\subsection{Estimating the Number of Active User}

First, we show how to estimate the number of active users based on the perfect covariance matrix estimate $\hat{\boldsymbol \Sigma}_{\boldsymbol Y}(\lambda_1,...,\lambda_N)$.  With pilot matrix $\boldsymbol A (\boldsymbol \phi)$, the received measurement covariance matrix  ${\boldsymbol \Sigma}_{\boldsymbol Y}(\lambda_1,...,\lambda_N)$ given in (4) can be re-expressed as 
\begin{align}
\boldsymbol \Sigma_{\boldsymbol Y}(\lambda_1,...,\lambda_N)=  \bar{\boldsymbol A}(\boldsymbol \phi) \bar{ \boldsymbol \Gamma} \bar{\boldsymbol A}^H(\boldsymbol \phi)+ \sigma^2 \boldsymbol I_L =  \hat{\boldsymbol \Sigma}_{\boldsymbol Y}(\lambda_1,...,\lambda_N),
\end{align}
where $ \bar{\boldsymbol A}(\boldsymbol \phi) = [..., \boldsymbol a(\phi_k),...], \forall k\in \mathcal{K}$, and $\bar{ \boldsymbol \Gamma} = \mathrm{diag}(..., \beta_k, ...), \forall k\in \mathcal{K}$. Since the columns in the pilot matrix $ \bar{\boldsymbol A}(\boldsymbol \phi)$ are linearly independent according to  {Theorem 1}, it follows that the rank of the matrix  
\begin{align}
\boldsymbol Q = \hat{\boldsymbol \Sigma}_{\boldsymbol Y}(\lambda_1,...,\lambda_N) - \sigma^2 \boldsymbol I_L = \bar{\boldsymbol A}(\boldsymbol \phi) \bar{ \boldsymbol \Gamma} \bar{\boldsymbol A}^H(\boldsymbol \phi)
\end{align}
is just  the number of active user $K$, i.e., $K = \mathrm{rank}(\boldsymbol Q)$. As a result, if the covariance matrix estimate  $\hat{\boldsymbol \Sigma}_{\boldsymbol Y}(\lambda_1,...,\lambda_N)$  is perfect, the number of active users can be accurately estimated.
\vspace{-2.5em}

\subsection{Active User Detection}
After detecting the number of active users, we show how to detect the active users based on the covariance matrix  $\hat{\boldsymbol \Sigma}_{\boldsymbol Y}(\lambda_1,...,\lambda_N)$.  Since each pilot sequence  $\boldsymbol a(\phi_k)$ is determined by the signature parameter  $\phi_k$,  detecting the active users is equivalent to finding parameters $\{\phi_k\}_{k=1}^K$  from the covariance matrix $\hat{\boldsymbol \Sigma}_{\boldsymbol Y}(\lambda_1,...,\lambda_N)$. To detect $\phi_k$'s, we first have the following property
\begin{align}
\boldsymbol A_1(\boldsymbol \phi) = \boldsymbol A_0(\boldsymbol \phi) \boldsymbol \Phi ,
\end{align}
where $\boldsymbol A_0(\boldsymbol \phi) \in \mathbb C^{(L-1)\times K}$ collects the first row to the second last row of the pilot matrix $\bar{\boldsymbol A}(\boldsymbol \phi)$, and $\boldsymbol A_1(\boldsymbol \phi) \in \mathbb C^{(L-1)\times  K}$ collects the second row to the last row of the pilot matrix $\bar{\boldsymbol A}(\boldsymbol \phi)$, i.e., 
\begin{align}
&\boldsymbol A_0(\boldsymbol \phi)  = \left[
 \begin{matrix}
   &\boldsymbol a(\phi_1)(1) & ... & \boldsymbol a(\phi_K)(1) \\
       &  \vdots  & \vdots  & \vdots  \\
    &\boldsymbol a(\phi_1)(L-1) & ... & \boldsymbol a(\phi_K)(L-1) 
  \end{matrix}
  \right], \\
  &\boldsymbol A_1(\boldsymbol \phi)  = \left[
 \begin{matrix}
   \boldsymbol a(\phi_1)(2) & ... & \boldsymbol a(\phi_K)(2) \\
         \vdots & \vdots & \vdots \\
         \boldsymbol a(\phi_1)(L) & ... & \boldsymbol a(\phi_K)(L) 
  \end{matrix}
  \right],
  \end{align}
and 
\begin{align}
\boldsymbol \Phi  = &\mathrm{diag}\{  \exp(-j2\pi \delta \cos \phi_1 /\lambda),  \exp(-j2\pi \delta \cos \phi_2 /\lambda), \nonumber \\
&..., \exp(-j2\pi \delta \cos \phi_{K} /\lambda)\}. 
\end{align}

If $\boldsymbol A_0(\boldsymbol \phi)$ and  $\boldsymbol A_1(\boldsymbol \phi)$ are known, we could obtain $\boldsymbol \Phi $ by solving (10). However, they are unknown in general. In the following, we show that $\boldsymbol \Phi$ still can be efficiently obtained based on $\boldsymbol Q$, which is computed from $\hat{\boldsymbol \Sigma}_{\boldsymbol Y}(\lambda_1,...,\lambda_N)$ using (9). Specifically, the EVD of $\boldsymbol Q$ is defined as  
\begin{align}
\boldsymbol Q = \boldsymbol U \boldsymbol S \boldsymbol U^H,
\end{align}
where $\boldsymbol U=[\boldsymbol u_1,..., \boldsymbol u_L]$ consists of the eigenvectors of $\boldsymbol Q$, and $\boldsymbol S= \mathrm{diag}\{s_1,...,s_K,0,...,0\}$ consists of the eigenvalues of $\boldsymbol Q$  (only $K$ of them are non-zero). Then, we have the following lemma.

\begin{lemma}
There exists an invertible matrix $\boldsymbol C \in \mathbb C^{K \times K}$ such that 
\begin{align}
\bar{\boldsymbol U}  = \bar{\boldsymbol A}(\boldsymbol \phi) \boldsymbol C,
\end{align}
where matrix $\bar{\boldsymbol U} = [ \boldsymbol u_1,  ...,  \boldsymbol u_k]$.
\end{lemma}

A rigorous proof of Lemma 1 can be found at [8, page 285].  Similar to the construction of $\boldsymbol A_0(\boldsymbol \phi)$ and $\boldsymbol A_1(\boldsymbol \phi)$ based on $\bar{\boldsymbol A}(\boldsymbol \phi)$,  define   $ \bar{\boldsymbol U}_0$ and $\bar{\boldsymbol U}_1$  as
\begin{align}
& \bar{\boldsymbol U}_0 = \left[
 \begin{matrix}
   &\boldsymbol u_1(1) & ... & \boldsymbol u_K(1) \\
        &\vdots & \vdots & \vdots \\
         &\boldsymbol u_1(L-1) & ... & \boldsymbol u_K(L-1) 
  \end{matrix}
  \right],    \\
&\bar{\boldsymbol U}_1  = \left[
 \begin{matrix}
   &\boldsymbol u_1(2) & ... & \boldsymbol u_K(2) \\
       &  \vdots  & \vdots  & \vdots  \\
    &\boldsymbol u_K(L) & ... & \boldsymbol u_K(L) 
  \end{matrix}
  \right].
  \end{align}
It can be shown that 
\begin{align}
&\boldsymbol \bar{\boldsymbol U}_1 =  \boldsymbol A_1(\boldsymbol \phi)  \boldsymbol C =  \boldsymbol A_0(\boldsymbol \phi) \boldsymbol \Phi \boldsymbol C \nonumber \\
&= \boldsymbol A_0(\boldsymbol \phi) \boldsymbol C \underbrace{ \boldsymbol C^{-1}  \boldsymbol \Phi \boldsymbol C }_{\triangleq  \boldsymbol \Psi} = \boldsymbol \bar{\boldsymbol U}_0 \boldsymbol \Psi.
\end{align}
Equation (18) reveals an efficient way to obtain $\boldsymbol  \Phi$. First, it can be shown that the columns in $\boldsymbol \bar{\boldsymbol U}_0$  are linearly independent with each other under our interested regime of $L>K$. As a result, $\boldsymbol \Psi$ can be uniquely obtained by 
\begin{align}
\boldsymbol \Psi = \left[\boldsymbol \bar{\boldsymbol U}_0^H \boldsymbol \bar{\boldsymbol U}_0 \right]^{-1} \boldsymbol \bar{\boldsymbol U}_0^H \boldsymbol \bar{\boldsymbol U}_1.
\end{align}
Next, it can be observed in (18) that the EVD of $\boldsymbol \Psi$ is defined by 
\begin{align}
\boldsymbol \Psi=  \boldsymbol C^{-1}  \boldsymbol \Phi \boldsymbol C.
\end{align}
As a result, we can compute the EVD of $\boldsymbol \Psi$ to obtain $\boldsymbol \Phi$ accurately. Define $\boldsymbol \Phi(k)$ as the $k^{th}$ diagonal element of $\boldsymbol \Phi$. According to (13), $\phi_k$ can be obtained by
\begin{align}
\phi_k =  \arccos \left( \frac{-j \ln \boldsymbol \Phi(k) }{2\pi \delta}\right), k = 1 ,..., K.
\end{align}
Since the BS knows the signature parameters $\{\phi_n\}_{n=1}^N$ for all the users, the active users can be accurately detected by the BS. 

To summarize, we have the following theorem.

\begin{theorem}
Given perfect estimation of $\boldsymbol \Sigma_Y (\lambda_1,...,\lambda_N)$ in the asymptotic regime of  $M \rightarrow \infty$, and under the pilot design given in (7), the active users can be always perfectly detected based on the following algorithm (Algorithm 1) when $L> K$.
\end{theorem}
\begin{algorithm}[!h]
\caption{The proposed scheme when  $M \rightarrow \infty$}
\noindent {\bf Users:} Each user generates its pilot vector $\boldsymbol a (\phi_n)$ using (7).
\noindent {\bf Base Station:}  

\noindent {\bf 1:}  Compute the measurement covariance matrix estimate  $ \hat{\boldsymbol \Sigma}_{\boldsymbol Y} (\lambda_1,...,\lambda_N)$ using (5);

\noindent {\bf 2:} Compute matrix $\boldsymbol Q$ using (9) and then recover the active user number  $K = \mathrm{rank}(\boldsymbol  Q)$.

\noindent {\bf 3:} Find eigenvectors $\boldsymbol U = [\boldsymbol u_1, ... \boldsymbol u_L]$ from EVD of $\boldsymbol Q$ using (14). Let  $\bar{\boldsymbol U} = [\boldsymbol u_1,..., \boldsymbol u_K]$.Then construct matrix $\bar{\boldsymbol U}_0$   and   $\bar{\boldsymbol U}_1$  based on (16) and (17). 

\noindent {\bf 4:} Compute matrix $\boldsymbol \Psi$ using (19), based on which matrix  $\boldsymbol \Phi$ is computed from the EVD of $\boldsymbol \Psi$ in (20).  Then, the pilot signature parameters $\{\phi_k\}_{k=1}^K$ are obtained via (21).

\noindent {\bf 5:}  After matching parameters $\{\phi_k\}_{k=1}^{K}$ to the signature $\{\phi_n \}_{n=1}^N$ at the BS, the active set $\mathcal K$ is obtained. 
\end{algorithm}

\vspace{-18pt}
\subsection{Channel Estimation}

After detecting the active users, the received data model  (3) reduces to 
\begin{align}
\boldsymbol Y=\bar{\boldsymbol A}(\boldsymbol \phi)  \bar{\boldsymbol \Gamma} \bar{\boldsymbol H} + \boldsymbol Z,
\end{align}
where $\bar{\boldsymbol H}=[..., \boldsymbol h_k, ... ]~\forall k \in \mathcal K$, and $\bar{\boldsymbol A}(\boldsymbol \phi)$ is known after $\{\phi_k\}_{k=1}^K$ are detected perfectly. Then the MMSE estimator for channel estimation can be expressed as \cite{Kay}:
\begin{align}
\hat{\boldsymbol H} =   \bar{\boldsymbol \Gamma}\bar{\boldsymbol A}(\boldsymbol \phi)^H \left( \bar{\boldsymbol A}(\boldsymbol \phi)    \bar{\boldsymbol \Gamma}   \bar{\boldsymbol A}(\boldsymbol \phi)^H + \sigma^2 \boldsymbol I  \right)^{-1} \boldsymbol Y, 
\end{align}
with the mean-square-error (MSE) as \cite{Kay}:
\begin{align}
\boldsymbol \Sigma_{\hat{\boldsymbol H}}  \!= \! \bar{\boldsymbol \Gamma}\! -\! \bar{\boldsymbol \Gamma}\bar{\boldsymbol A}(\boldsymbol \phi)^H\left(  \bar{\boldsymbol A}(\boldsymbol \phi) \bar{\boldsymbol \Gamma} \bar{\boldsymbol A}(\boldsymbol \phi)^H \!+\! \sigma^2 \boldsymbol I    \right)^{-1}    \bar{\boldsymbol A}(\boldsymbol \phi) \bar{\boldsymbol \Gamma}.
\end{align}

\vspace{-12pt}
\section{Algorithm Design with a Large but Finite $M$}
In a practical massive MIMO system where the number of antennas at the BS $M$ is large but  finite, the estimation of the covariance matrix $\boldsymbol \Sigma_Y (\lambda_1,...,\lambda_N) $ based on (5) is not perfect, i.e., $\hat{\boldsymbol \Sigma}_{\boldsymbol Y} (\lambda_1,...,\lambda_N)  \approx \boldsymbol \Sigma_Y (\lambda_1,...,\lambda_N)$. The estimation error will degrade the user detection performance of our proposed scheme in the following manners. First, since $\boldsymbol Q \neq \bar{\boldsymbol A}(\boldsymbol \phi) \bar{ \boldsymbol \Gamma} \bar{\boldsymbol A}^H(\boldsymbol \phi) $,  the rank of $\boldsymbol Q$ may not be equal to the rank of $\bar{\boldsymbol A}(\boldsymbol \phi) \bar{ \boldsymbol \Gamma} \bar{\boldsymbol A}^H(\boldsymbol \phi)$, leading to an error in the estimation of the number of active users.  Second, due to the estimation error, 
there may not exist an invertable $\boldsymbol C$ such that (15) holds. As a result, (18) is no longer true and the estimation of $\phi_k$'s based on (19)-(21) is not perfect. 
%
%

To tackle the first difficulty, we still calculate the EVD of $\boldsymbol Q$ and obtain its eigenvalues $\{s_l  \}_{l=1}^L$. Next, the number of active users  $\hat{K}$ can be estimated as the number of eigenvalues that are sufficiently  larger than a threshold $\eta > 0$, i.e., 
\begin{align}
\hat K = |  \{s_l :  s_l > \eta , l = 1,  ...,L \} |.
\end{align}

For the second difficulty, the signature parameter can still be obtained from (21), which are denoted by  $\{\hat{\phi}_k\}_{k=1}^{\hat K}$. Due to the estimation error, it is in general hard to find a value of $\phi_k$ that is exactly equal to the estimated $\hat{\phi}_k$. In this case, we claim that one user $k^a$ is active if its $\phi_{k^a}$ is closest to a particular $\hat{\phi}_k$, i.e.,
\begin{align}
\hat{\mathcal K}  = \{k^a: {\arg\min}_{n=1,...,N}  (\phi_n - \hat{\phi}_k)^2 , k=1,...,K\}.
\end{align}
In summary, the proposed covariance-based algorithm for user activity detection in the case with a large but finite $M$ is presented in {Algorithm 2}. Then, the channels of the detected users can be estimated based on (23).

\begin{algorithm}[!h]
    \caption{The proposed scheme with a large but finite $M$}
\noindent {\bf Users:} Each user generates its pilot vector $\boldsymbol a (\phi_n)$ using (7).
\noindent {\bf Base Station:}  

\noindent {\bf 1:}  Compute the measurement covariance matrix estimate  $ \hat{\boldsymbol \Sigma}_{\boldsymbol Y} (\lambda_1,...,\lambda_N)$ using (5);

\noindent {\bf 2:} Compute matrix $\boldsymbol Q$ using (9) and its eigenvalues $\{s_l \}_{l=1}^L$ using (14), based on which the active user number is estimated using (25).

\noindent {\bf 3:} Find eigenvectors $\boldsymbol U = [\boldsymbol u_1, ... \boldsymbol u_L]$ from EVD of $\boldsymbol Q$ using (14). Let  $\bar{\boldsymbol U} = [\boldsymbol u_1,..., \boldsymbol u_K]$.Then construct matrix $\bar{\boldsymbol U}_0$   and   $\bar{\boldsymbol U}_1$  based on (16) and (17). 

\noindent {\bf 4:} Compute matrix $\boldsymbol \Psi$ using (19), based on which matrix  $\boldsymbol \Phi$ is computed from the EVD of $\boldsymbol \Psi$ in (20).  Then, the pilot signature parameters $\{\phi_k\}_{k=1}^K$ are obtained via (21).

\noindent {\bf 5:}  Estimate  the active set $\hat{\mathcal K}$ using (26). 
\end{algorithm}

\section {Comparison with Existing Schemes}
In the literature, AMP was first proposed in \cite{Liang_TSP} to detect the active users and estimate their channels. Then, a covariance-based approach was proposed in \cite{Con1, Con2} to merely detect the active users. In this section, we compare our proposed scheme with the existing schemes in terms of design philosophy, performance, and complexity.


\subsubsection{Design Philosophy} AMP views user detection and channel estimation as a compressed sensing problem based on the observation that $\boldsymbol X$ is row-sparse. The idea is to recover $\boldsymbol X$ based on (3). On the other hand,  the covariance-based method proposed in \cite{Con1, Con2} aims to recover $\boldsymbol \Gamma$ based on (4). Different from these two existing  schemes,  we construct the pilot sequences in a way that each $\phi_k$ is the identifier of each user $k$. This enables an efficient algorithm to estimate the number of active users and detect the active users.

\subsubsection{Performance}  It can be shown that in the asymptotic regime of $M \rightarrow \infty$, all the AMP algorithm \cite{Liang_TSP}, the covariance-based algorithm proposed in \cite{Con1, Con2}, and our proposed algorithm can achieve perfect user activity detection. Particularly, AMP requires  $L$ to be proportional to $K$, the covariance-based algorithm \cite{Con1, Con2} requires $L$ to be linear to $\sqrt{K}$, and our proposed scheme requires $L>K$. Although the covariance-based algorithm \cite{Con1, Con2}  requires the minimum $L$ for user detection, the algorithm cannot be applied to estimate the channels since $L$ is much smaller than $K$. However, our proposed scheme can estimate the channels as well. 


In the practice regime with large but finite $M$, all the algorithms cannot achieve perfect user activity detection. Particularly, if $N$ is too large, the minimum gap between $\phi_n$'s, i.e., $\pi/N$, is too small under our proposed scheme, In this case, a small error in the estimation of $\phi_k$ may lead to a missed detection event for an active user $k$. As a result, as will be shown in the numerical examples in Section VI, our proposed scheme works quite well when the number of users is moderate, e.g.,  up to about 100 users. 

\subsubsection{Computational Complexity} From Algorithm 2, it is observed that the computational complexity is dominated by two EVD computations in Step 3 and 4 and one pseudo-inverse computation in Step 4, resulting in the complexity order of $O(L^3 + K^3 + K^2(L-1)+L^2M)$. Due to the sporadic nature of the wireless traffic and the ultra-latency requirement, the active user number $K$ and the pilot length $L$ are usually not very large. Furthermore, the proposed scheme is a closed-form algorithm, with each step  only involving basic algebraic operations.   In contrast, both the AMP and covariance-based algorithms are iterative algorithms, with complexity order in each iteration  of  $O(LNM) $ and  $O(L^3)$ respectively.

\vspace{-10pt}
\section{Numerical results}



In this section, numerical results are presented to assess the performance of the proposed algorithm. The simulation setup is as follows. There are $N = 100$ users in the cell, and in each coherent block $K = 5$ users are active. The BS is equipped with $M = \{8, 16, 32, 64, 128 \}$ antennas and the pilot sequence length is $L = 12$. The path loss model of the wireless channel for user $n$ is given as $g_n = -128.1 - 36.7 \log_{10} (d_n)$ in dB, in which $d_n \in [0, 100m]$ denotes the distance between the BS and the $n^{th}$ user. The bandwidth of the wireless channel is $10$ KHz and the transmit power for each user is $25$ dBm. The power spectral density of the noise is -169 dBm/Hz.  All the simulation results are obtained by averaging over 10000 simulation trials.


First, for user activity detection, the proposed scheme is compared to the  the covariance-based algorithm \cite{Con1, Con2} \footnote{The AMP algorithm works in the regime of $K>M$. As a result, its performance cannot be shown in our considered setup.}. Fig.1 compares the probability of miss detection and false alarm between the two algorithms  under different number of antennas.  To conveniently show the detection error behavior of the covariance-based method, its detection threshold is properly set to achieve the same probability of miss detection of the proposed scheme.  It is observed from Fig. 1 that as the antenna number increases, the detection error probabilities of the proposed scheme decrease due to the better estimation of the covariance matrix. Moreover, it is observed  that with the same probability of miss detection, the proposed scheme achieves much smaller probability of false alarm than the covariance-based method proposed in \cite{Con1, Con2}. 

\begin{figure}[!h]
\centering
\vspace{-13pt}
\includegraphics[width= 3 in]{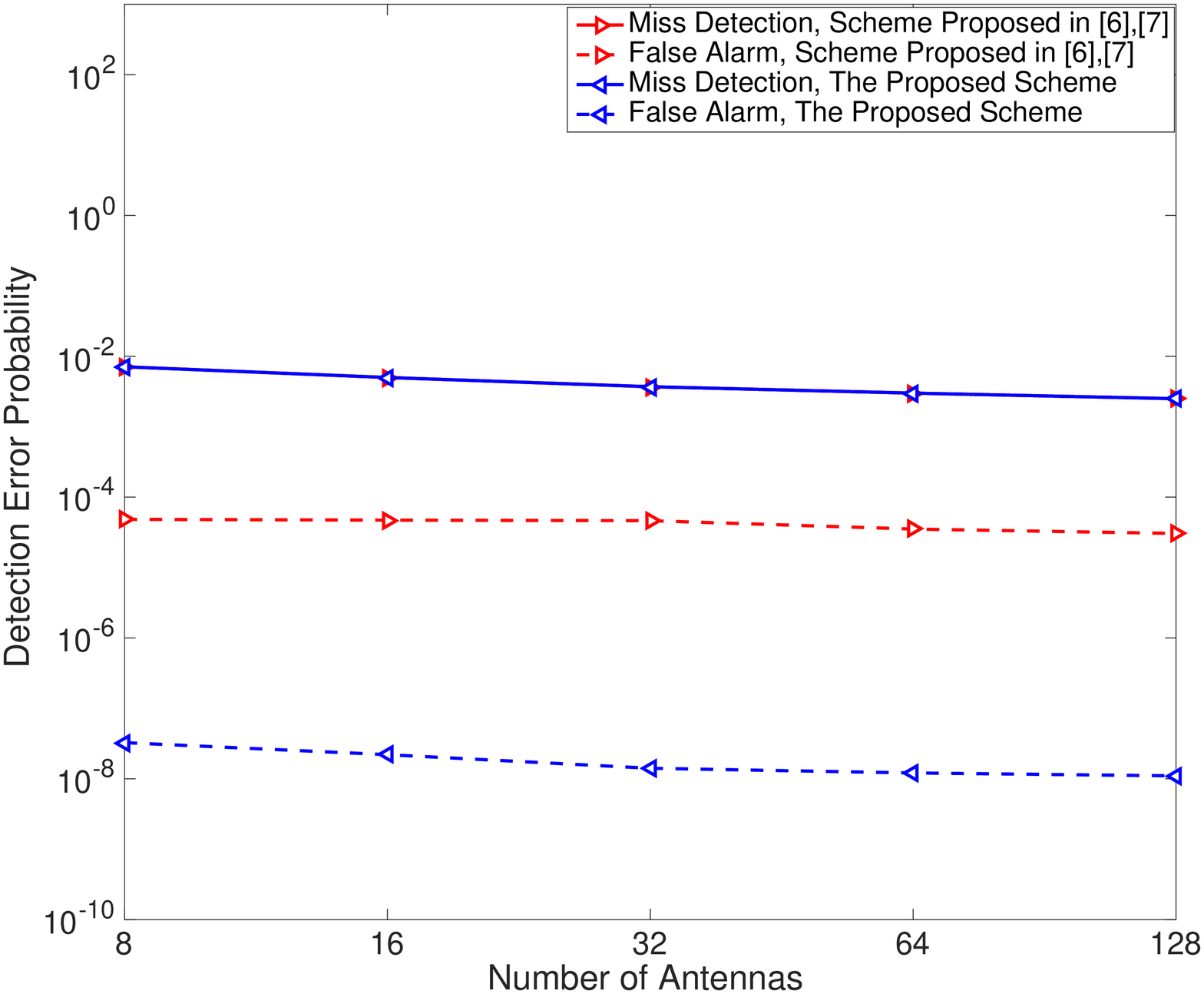}    
\caption*{Fig. 1. Performance of activity detection versus antenna numbers}
\vspace{-10pt}
\label{fig_topology}
\end{figure}


Moreover, the computational time of the proposed scheme and the covariance-based algorithm proposed in \cite{Con1, Con2} is given in Table I. It is observed that our proposed closed-form scheme is orders of magnitude faster than the iterative scheme proposed in \cite{Con1, Con2}.

\begin{table}[!t]
\centering    
\caption{Computational time (sec) versus different antenna numbers}
\begin{tabular}{|c|c|c|c|c|c|}
\hline
Number of antennas at BS                       & 32                    & 64       & 128             \\ \hline
The proposed scheme      &\! $4.91 \!\! \times \!\!10^{-4}$ \!& $5.48 \!\!\times \!\!10^{-4}$ & $6.01 \!\!\times\!\! 10^{-4}$  \\ \hline
Scheme proposed in [6],[7]                 & 0.0567                & 0.0573      &      0.0586    \\ \hline
\end{tabular}
\vspace{-13pt}
\end{table}

Next, Fig. 2 shows the channel estimation performance assuming a set of active users have been detected perfectly. For the benchmark scheme, we consider the case where each pilot symbol is i.i.d. Gaussian distributed as for the AMP algorithm. It is observed from Fig. 2 that under our pilot design (7) , the MSE for channel estimation is much lower than that for the benchmark scheme.

\begin{figure}[!h]
\centering
\vspace{-11pt}
\includegraphics[width= 3 in]{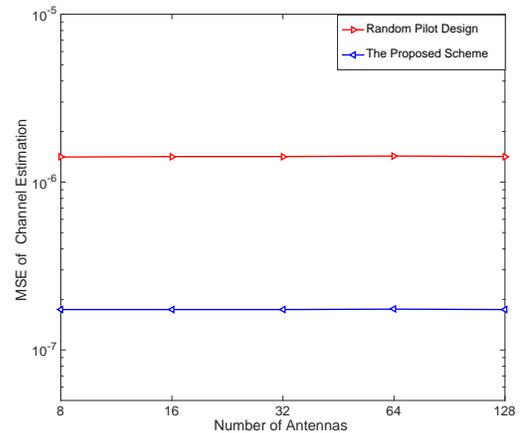}    
\caption*{Fig.2. Performance of channel estimation versus antenna numbers}
\vspace{-11pt}
\label{fig_topology}
\end{figure}

At last, it is worth mentioning that our scheme works in the regime of $M \gg K$ and $N$ is moderate. If both $K$ and $N$ are extremely large, AMP can yield better performance in this regime. 

\vspace{-13pt}
\section{Conclusions}
In this paper, a novel covariance-based scheme was proposed to detect the active users and then estimate their channels in mMTC, assuming no prior information about the active user number. By judiciously designing identifiable constant-modulus pilot signals, the proposed scheme was able to perform user activity detection in a closed-form manner, based on which the MMSE channel estimator is derived.

\end{document}